\documentclass[fleqn,twoside]{article}
\usepackage{espcrc2}
\usepackage{amssymb}

\usepackage{graphicx}
\usepackage[figuresright]{rotating}

\newcommand{\AmS}{{\protect\the\textfont2
  A\kern-.1667em\lower.5ex\hbox{M}\kern-.125emS}}

\title{IceCube and Searches for Astrophysical Sources}

\author{T. Montaruli\address[UW]{Physics Department, University of Wisconsin, Madison, WI-53706 } for the IceCube Collaboration%
}
       
\begin{document}

\begin{abstract}
Understanding cosmic acceleration mechanisms, such as jet formation in black holes, star collapses or binary mergers, and the propagation of accelerated particles in the universe is still a `work in progress' and requires a multi-messenger approach, exploiting the complementarities across all possible probes: ultra-high energy cosmic rays (UHECR), gamma-rays and neutrinos. In this report I will summarize some of the IceCube results concerning searches for astrophysical neutrino point sources and diffuse fluxes from 
populations of sources widely distributed in the sky or from the interactions of protons on the cosmic microwave background producing the GZK cut-off in the cosmic ray spectrum. I will compare the results to other neutrino telescopes  and to astrophysical models of neutrino production in sources. 
Another unresolved question concerns the nature of dark matter. Indirect searches have the opportunity to observe where it is located in the universe through the observation of secondary photons, neutrinos or antiparticles such as positrons and antiprotons. The potential for the search of neutrinos from the annihilation of WIMPs in IceCube is 
greatly enhanced by the addition of more compact strings, the DeepCore.
\vspace{1pc}
\end{abstract}

\maketitle

\section{INTRODUCTION}
\label{sec1}

\begin{figure}[t]
    \includegraphics[width=3.in,height=3.5in]{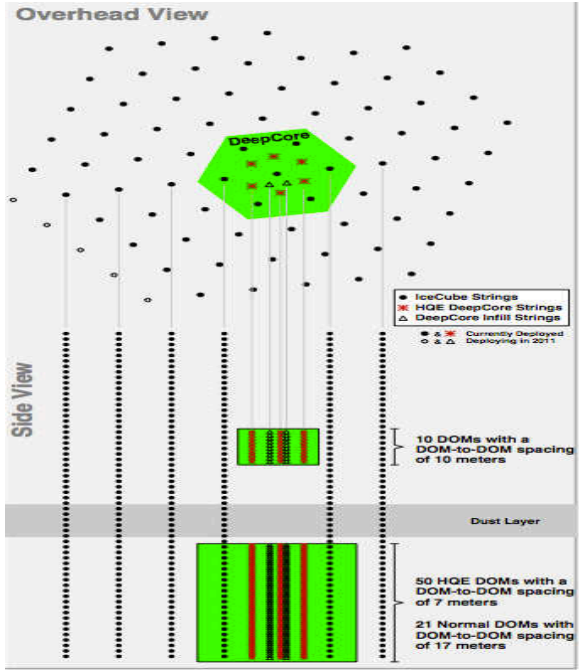}
 \caption{A schematic layout of IceCube DeepCore. The upper diagram shows a top view of the string positions in relation to current and future IceCube strings. It includes two additional strings, situated close to the central DeepCore string. The lower diagram shows the instrumented DeepCore region with the surrounding IceCube strings. \label{fig:fig1}}
\end{figure}

\begin{figure}[t]
     \includegraphics[width=3.in,height=3.in]{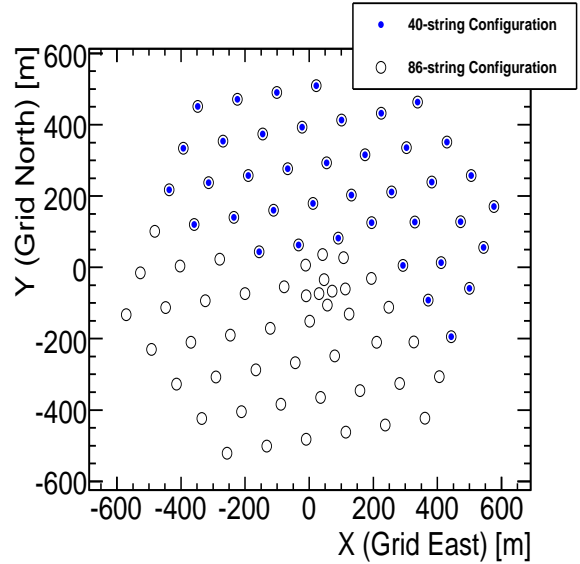}
\caption{\label{fig:fig2} Overhead view of the 40-string configuration, along with the additional strings that will make up the complete IceCube detector. }
\end{figure}

\begin{figure}[ht]
     \includegraphics[width=3.in,height=2.in]{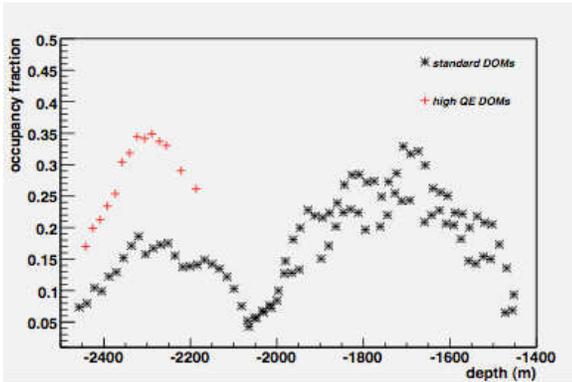}
\caption{\label{fig:fig3} In situ measurement of the occupancy versus depth. The lower black stars are for DOMs with standard PMTs, the upper red crosses are for HQE-DOMs.}
\end{figure}

The IceCube Neutrino Observatory will be composed of a deep array of 86 strings holding 5,160 digital optical modules (DOMs) deployed between 1.45 and 2.45 km below the surface of the South Pole ice and a surface array, IceTop, for extensive air shower measurements on the composition and spectrum of cosmic rays (CRs).
The strings are typically separated by about 125 m with DOMs vertically apart by about 17 m along each string.  
A DOM is a spherical, pressure-resistant glass vessel, housing a Hamamatsu R7081-02, 252~mm-diameter phototube (PMT) \cite{PMT}.
IceCube construction started with a first string installed in the 2005--6 season and will be completed in the austral summer of 2010--11. 
The running configuration consists of 79 strings. Six of these strings hold DOMs with higher quantum efficiency (HQE-DOMs). Laboratory measurements of the HQE-DOM quantum efficiency relative to a standard reference 2-inch PMT indicate values around $\sim 30\%$ compared to the standard DOMs with
$\sim 20\%$ at about 390 nm \cite{deepcorepaper}. The HQE-DOMs are located at smaller spacing of about 70 m horizontally and 7 m vertically in the layout shown in Fig.~\ref{fig:fig1}.
Together with two additional strings that will be deployed during the last construction season and seven standard strings of IceCube
they make up DeepCore, designed to enhance the physics performance of IceCube below 1 TeV such as dark matter searches.  
Fig.~\ref{fig:fig2} shows the layout of the 86 strings of the full detector and of the 40-strings configuration, that
was used for most of the analysis results I will summarize here.  Fig.~\ref{fig:fig3} shows the occupancy (the fraction of events with a DOM that detected at least 1 photon on a string where at least seven other DOMs have also detected photons) versus depth \cite{deepcorepaper}. In a uniform medium, such a plot would exhibit a decreasing trend due to muon energy losses in the ice, but in IceCube it reflects the convolution between energy losses and ice depth dependent optical properties.
As a matter of fact, the effect of a big dust layer, most probably with a component of volcanic nature, is visible between -2,150 and -1,950 m.
Moreover, the deep ice is so transparent that light propagates up to $\sim 600$~m and the hit rate is higher than it would be in a uniform medium.
The plot explains why no DeepCore DOM is installed in the dust region. The plot also shows the increase in efficiency of the HQE-DOMs at the expense of some increase of the dark noise rate (from about 500 Hz for normal DOMs to about 700 Hz in HQE-DOMs). 

The charge and time of Cherenkov photons induced by relativistic charged particles passing through the ice sheet is detected and the PMT signal is digitized with dedicated electronics included in the DOMs \cite{WF}. The DAQ hardware is made of two waveform digitization systems using Analog Transient Waveform Digitizers (ATWD) that record the details of the first 400 ns of the waveform at 300 MSPS and fADCs (fast ADC) using a 40 MSPS commercial ADC chip
covering up to 6.4 $\mu$s. A digitization cycle is initiated by a discriminator trigger with threshold set at a voltage corresponding to about 1/4 photoelectron. When this happens, the FPGA starts ATWD and fADC digitization on the next clock edge and  data are sent to the surface when a coincidence of at least one other hit in the nearest or next-to-nearest neighboring DOMs within $\pm 1$~$\mu$s is satisfied (Hard Local Coincidence - HLC). After the 40-string run, Soft Local Coincidences have been added to the acquisition and then to reconstructions. In this case even isolated hits are stored saving only fADC information.

Various triggers are used in IceCube. Most of the results shown here are based on a 
simple multiplicity trigger requiring that the sum of all HLC hits in a rolling time window of $5\,\mu$s is
above 8 (SMT8). The duration of the trigger is the amount of time that this counter stays at or above 8 as the time window keeps
moving.  
Once the trigger condition is met, all local coincidence hits are read in a readout window of $\pm 10\,\mu$s for the 40-string run and 
of $^{+6}_{-4} $ $\mu$s (to reduce the noise rate) in the 79-string run.   
IceCube triggers primarily on down-going muons at a rate of about 950~Hz in the 40-string configuration and about 1.8 kHz in the 79-string one.  
Variations in the trigger rate by about $\pm10\%$ are due to seasonal changes. 
During the austral summer the atmosphere above the South Pole gets thinner and the probability of pions generated in the CR induced cascades to decay rather than interact increases - and hence the muon rate \cite{Tilav:2010hj}.  

\begin{figure}[ht]
     \includegraphics[width=3.2in,height=2.2in]{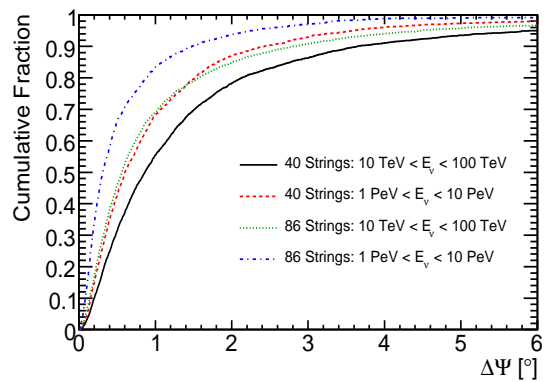}
\caption{\label{fig:fig4}Point spread function (fraction of events included in the angle shown on the x axis between the neutrino simulated direction and the reconstructed muon
one) for the 40 and 86-string configurations and for two energy bins at the final level of cuts for the point source search \protect\cite{jon}.}
 \end{figure}

Direction of events can be reconstructed using the time of hit PMTs and the amplitude as well, and the energy can be inferred exploiting the stochastic energy loss
properties of muons and the charge measurement. The muon energy resolution on an event by event basis is limited (about an uncertainty of a factor of 2 on the muon energy 
at the core of the detector between $10^4 -- 10^8$~GeV) but the capability of reconstructing spectra (as shown in Sec.~\ref{sec2}) is helped by the wide lever arm of many orders of magnitude in energy to which IceCube is sensitive. The point spread function is shown in Fig.~\ref{fig:fig4}. It is determined by the kinematic angle between the neutrino and the reconstructed secondary muon and by the intrinsic angular resolution of the detector (limited mainly by scattering of photons in the ice and by the PMT transit time spread).
Two bins in energy are shown: for higher energies tracks are longer across the array and can be better reconstructed. The high energy muons produced by muon neutrino interactions point back to the neutrino source with degree-level accuracy making IceCube a `neutrino telescope'. 
Neutrino events would cluster around the point source producing them within an uncertainty of less than $1^o$ for 50\% of the events with $E_\nu \in [10,100]$ TeV 
and within $0.6^o$ for 50\% of the events with $E_{\nu} \in [1,10]$ PeV for 40 strings.
The absolute pointing accuracy has been confirmed studying a deficit of muons due to cosmic rays blocked by the Moon disc.  The Moon shadow detection, initially reported in \cite{Moon}, is seen at the level of 6.76$\sigma$ for 14 lunar cycles with the 40-string configuration.
New data increase this significance.
IceCube is also sensitive to shower events induced by neutral current and electron/tau charged current interactions. These events have no pointing capability but the energy can be reconstructed.

\begin{figure} [t]
\includegraphics[width=3.2in,height=2.5in]{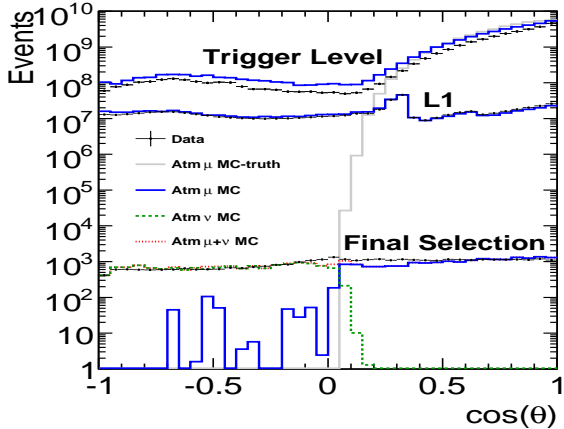}
\caption{\label{fig:filtering} Distribution of the reconstructed cosine zenith of tracks at trigger level, L1, and final cut level for point source analysis for data and simulation of atmospheric muons (using the composition model in \cite{Hoerandel:2002yg}) and atmospheric neutrinos \cite{Barr:2004br}. The deficit of Monte Carlo (MC) horizontal events compared to data is most likely caused by a deficiency in knowledge of the CR composition in the region around the knee that would require a harder component of protons or of heavier elements \protect\cite{jon}.}
\end{figure}
In this report, I will focus on results for astrophysical sources of neutrinos and on dark matter searches. Many results concern the 
40-string configuration in operation from 2008 April 5 to 2009 May 20.  
Over the entire period the detector ran with an uptime of 92\%, yielding 375.5 days of total exposure.  
Deadtime is mainly due to test runs during and after the construction season dedicated to calibrating the additional strings and upgrading data acquisition systems.  
The data and MC comparison is summarized in Fig.~\ref{fig:filtering} for different levels: trigger level, online filtering (L1) that runs at the South Pole
 with very basic reconstruction and cuts that reduce for 40-strings the trigger rate of about 1~kHz to 22~Hz and final level of cuts for the point source analysis \cite{jon}.
 These cuts are based on variables such as the quality and the angular error of the reconstruction, the number of hits due to unscattered Cherenkov light from the track
 and the total charge detected in one event.
 Another intriguing subject, covered in another report in these proceedings \cite{simona}, is the observation of anisotropies at large ($\gtrsim 60^o$) and intermediate scale ($\sim 20^o$) in the muon flux produced by cosmic ray (CR) interactions on the atmosphere.
\begin{figure}[htb]
\includegraphics[width=3.in,height=2.8in]{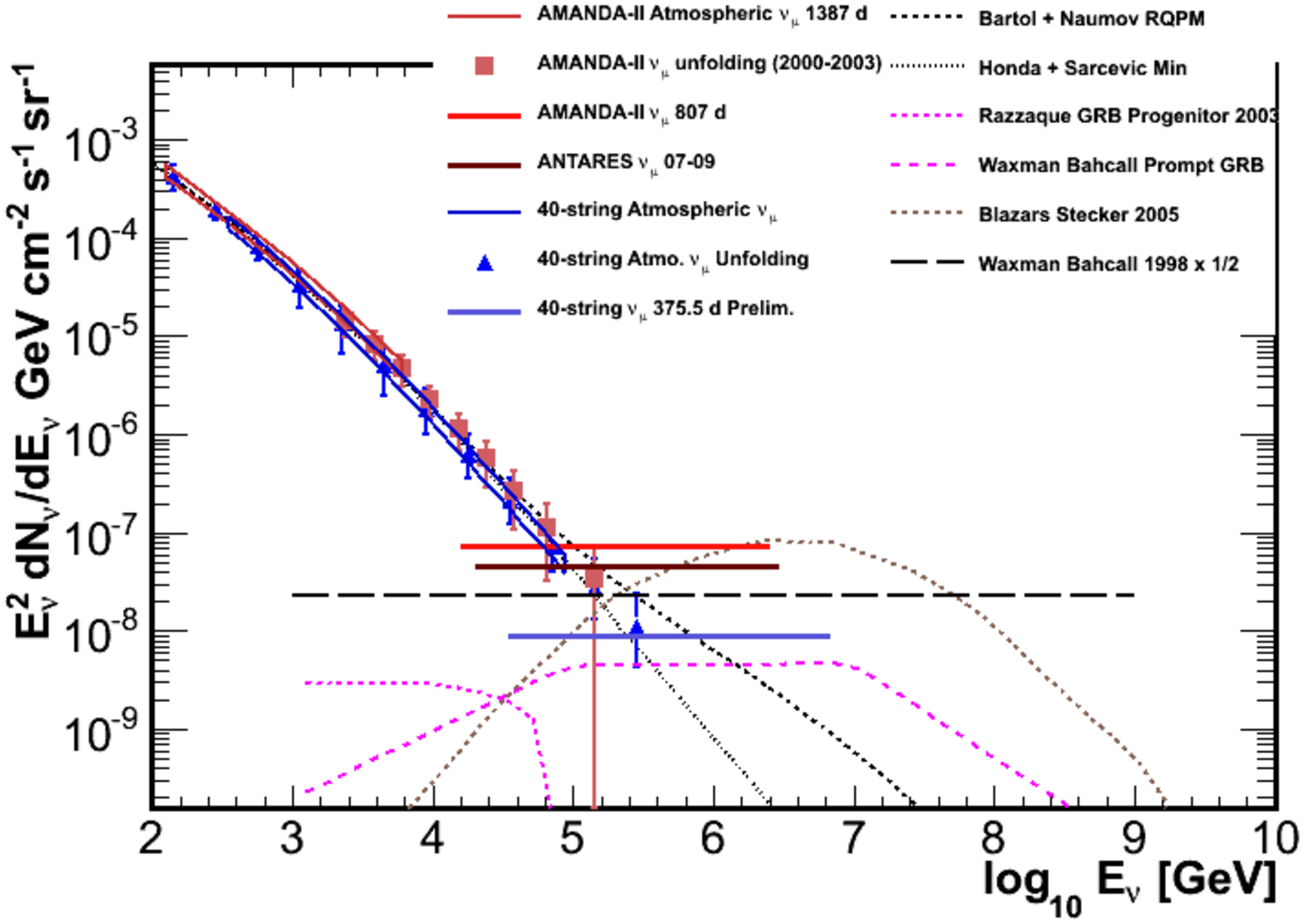} 
\caption{The flux of atmospheric $\nu_\mu+\bar{\nu}_{\mu}$ measured by AMANDA-II (squares) \protect\cite{julia} and 2 analyses with 40 strings of IceCube, an unfolding (triangles) \protect\cite{Warren} and a forward folding \protect\cite{Sean} (preliminary region between lines showing the range of uncertainty) is compared to atmospheric neutrino calculations
(high line \protect\cite{Honda:2006qj}; low line  \protect\cite{Barr:2004br}).
Horizontal lines are 90\% c.l. upper limit to an $E^{-2}$ muon neutrino flux for AMANDA-II \cite{julia} (807 d), ANTARES (334 d with 9-12 line configurations)
\protect\cite{Spurio} and 40 strings of IceCube (375.5 d). The WB limit \protect\cite{Waxman:1998yy} is shown together with some models on GRBs \protect\cite{Waxman:1998yy,Razzaque} and an example of an AGN model \cite{Stecker} rejected at 5$\sigma$ c.l. by the 40-string limit. }\label{fig:diffuse}
\end{figure}

\section{ASTROPHYSICAL NEUTRINO DIFFUSE FLUXES IN THE TEV-PEV and PEV-EHE ENERGY RANGES}
\label{sec2}

Unlike gamma and proton astronomy, neutrino astronomy can access the entire universe, probe cosmological sources and sources opaque to photons. 
Neutrino and proton astronomy are still not established since no 5-$\sigma$ level evidence of a signal has yet been observed, while TeV gamma astronomy is entering the time in which astronomical information is enabling physics interpretation. Proton astronomy is limited to the observation of less than about 100~Mpc by interactions of protons on the microwave background that above about $5 \times 10^{19}$~eV result in delta-resonance production. The observation of the drop off of the spectrum of Ultra-High Energy CR (UHECRs) above these energies is compatible with such a mechanism \cite{GZK}. The results on the composition are less clear and hint to a heavier component than pure protons in the UHECR flux. The minimum energy at which proton astronomy is possible is highly uncertain due to the incapability of calculating exactly proton deflections in the intergalactic and galactic magnetic fields. If UHECRs are heavy nuclei dominated, probably Fe deflections would be too high to allow astronomy at all.
One way to understand the UHECR composition is to look for neutrinos in coincidence with UHECRs possibly coming from common sources. 
In fact, if such correlation would be observed the composition must be proton dominated.
The correlation between the 22-string IceCube neutrino events and the highest energy events in Auger and HiRes used for astronomy studies has been investigated in \cite{lauer} providing a p-value of 1\% to occur as a random fluctuation of the background. The search is being extended to 82 UHECR detected by Auger and HiRes and the whole sky data sample of 36,900 muon events of 40-strings of IceCube described in Sec.~\ref{sec3}. The discovery potential at 5-$\sigma$ c.l. for 50\% of trials corresponds to 52 signal neutrinos in the direction of the UHECR. The signal and background from the UHECR directions is stacked and a tolerance value of $3^o$ around these directions is included to account for possible unknown magnetic filed deflections of UHECRs.

While the main limitation for proton astronomy is the uncertainty on  magnetic fields and the proton interactions with photon backgrounds, 
the main challenges for neutrino astronomy are due to
the interactions of neutrinos and to the atmospheric neutrino and muon backgrounds. Weak interactions make neutrinos unique probes to observe the far away universe and 
the inner part of sources from which photons or protons cannot escape, but they also require cubic-km scale detectors and beyond. 
The generic neutrino and CR source can be envisaged as an engine that accelerates protons, and magnetic fields in the accelerated region
confine them. If protons attain sufficient energy, they can interact with radiation and produce neutrons that can escape the accelerating region without interacting. Subsequently, neutron decay can give rise to the observed cosmic ray flux, gamma rays, and yet unobserved neutrinos. These three conditions together define an optically thin source~\cite{Ahlers:2005sn}. 
From the measured UHECR flux above $10^{18}$~eV the power required for a population of sources to generate the energy density observed in CRs can be inferred. This hints to active galactic nuclei (AGNs) and gamma-ray bursts (GRBs) as best candidate sources for the UHECRs and so for neutrinos coming from their interactions. 
This reasoning leads to an upper limit on the possible production of neutrinos from optically thin sources, the Waxman \& Bahcall (WB) upper limit~\cite{Waxman:1998yy} shown in Fig.~\ref{fig:diffuse}. It is extrapolated to lower energies than the UHECR region 
assuming a proton injection spectrum of $E^{-2}$~\cite{Waxman:1998yy} resulting from $1^{st}$ order Fermi acceleration in shocks and $p-\gamma$ interactions that produce 
$\Delta$-resonance. The limit also assumes that magnetic fields in the universe do not affect the observed CR spectrum. It has been divided by a factor of 2 to account for neutrino oscillations. The upper bound applies as well to $p-nucleon$ interactions since it is likely that the energy fraction of protons transferred to pions is even less than for $p-\gamma$ \cite{Waxman:1998yy}. The WB limit could be much higher at lower energies than $10^{16}$ eV assuming a 10\% contribution from an extragalactic source of protons on top of the measured galactic component but experimental limits exclude this option. It should be noted that if the injected CR spectrum from extragalactic sources includes heavy nuclei, then the upper bound and the cosmogenic neutrino fluxes \cite{Allard} would be lower than for the assumed proton case \cite{Murase:2010gj}. 

Fig.~\ref{fig:diffuse} shows the preliminary upper limit to a diffuse flux of neutrinos from extragalactic sources for an $E^{-2}$ neutrino spectrum.  It can be seen that the limit probes the region below the WB upper limit. The limit can be calculated for the case of specific models producing different spectra.
It can as well severely constrain the parameter space of models of neutrino production in AGNs normalized to the observations of Auger assuming that the UHECRs come from
a class of nearby AGNs, as shown in Ref.~\cite{Arguelles}. For instance, the region of the spectral index of the power law injection spectrum of UHECRs between -2 and -2.3 is allowed, while -2.7 is excluded.
Fig.~\ref{fig:diffuse} shows two measurements of the atmospheric neutrino spectrum, one performed with an unfolding method \cite{Warren} and another using a forward folding method, assuming a power law spectrum with nuisance parameters: the spectral index and the normalization of the flux \cite{Sean}. These two measurements are compatible and
extend the reach of AMANDA \cite{julia}. The statistics at high energy is not high enough to detect the charm component of the atmospheric neutrino flux
dominated by kaon decay in the IceCube energy range. 
The atmospheric neutrinos, as well as the atmospheric muons, are used for calibration of IceCube but are also a source of interesting particle physics information.
Data show a good agreement with the MC calculations created for explaining the data of previous generation experiments such as Super-Kamiokande and MACRO below the 10 TeV region \cite{Barr:2004br,Honda:2006qj}.
Above these energies the analytical functions in \cite{analytical} are used normalizing them to the MC calculations.
\begin{figure}[h]
\includegraphics[width=0.5\textwidth,height=0.44\textwidth]{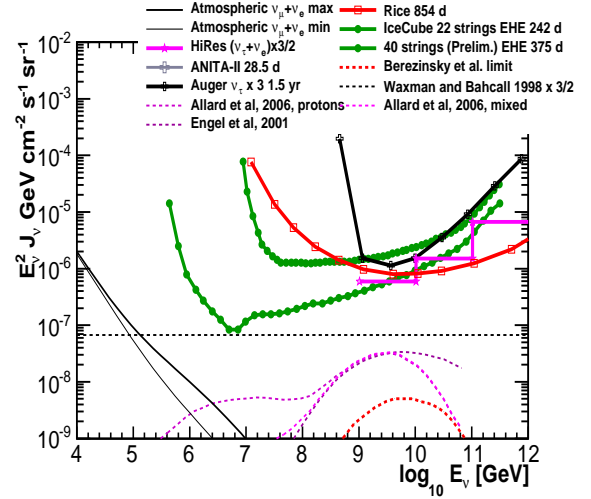}
\caption{90\%c.l. differential upper limits for experiments sensitive to all-$\nu$ flavors: Auger, HiRes, RICE \cite{Auger} and IceCube (the 22 string limit and the preliminary 40 string one \cite{Aya22}) are shown. For experiments sensitive to $< 3$ flavors the applied correction factor is given in the legend calculated under the assumption that neutrino fluxes at Earth are in a neutrino flavor ratio of 1:1:1 after oscillations.
At low energies the region of predictions of atmospheric $\nu_{e} + \nu_{\mu}$ is shown \protect\cite{Barr:2004br}. 
At higher energy cosmogenic neutrino models are shown \cite{Allard} as well as the limit on the cosmogenic neutrino flux calculated using the Fermi-LAT extragalactic $\gamma$-ray diffuse background limit \protect\cite{Berezinsky}. }\label{fig:ehe}
\end{figure}

When renouncing some tracking precision, the IceCube reach towards high energy increases, because PeV-EeV muons produce a large amount of
showers along the tracka large number of showers along the track which is a challenge for the reconstruction.
A search dedicated to EHE events, mainly using a zenith dependent cut on the total charge released in the detector,  has been performed using the data of 333.3 days of 22 strings of IceCube and 375.5 days of 40 strings \cite{Aya22}. Upper limits are shown in Fig.~\ref{fig:ehe}. 
Models predict between 2 and 24.5 (WB upper bound with z evolution of sources \cite{Waxman:1998yy}) neutrino events in 3 yrs of the full detector.

As discussed in \cite{Warren}, these diffuse flux searches are challenging since they require a prediction based on MC of expected fluxes and a 
very good description of the ice properties as a function of depth \cite{icepaper}, which recently has been considerably improved
using as calibration sources the flashers, LEDs included in the DOMs \cite{Dima}. Instrumental systematic errors are now estimated at the level of 20\% dominated by the ice description and the DOM efficiency. Another important factor under investigation is the effect of the CR composition and of hadronic models. 

\section{SEARCH FOR POINT SOURCES}
\label{sec3}

An unbinned likelihood (LH) method that compares the signal and the signal plus atmospheric muon and neutrino background hypotheses has been applied to look for emissions of neutrinos from point like sources. The method uses the reconstructed error on the direction and an energy proxy based on the estimate of the photon density along a muon track. Events would cluster around their source (see Fig.~\ref{fig:fig4}) and the neutrino signal from sources is expected to have a harder spectrum ($\sim E^{-2}$) compared to atmospheric muons and neutrinos produced in the meson decays in atmospheric showers ($\sim E^{-3.7}$ for $\gtrsim 500$~GeV).
The LH method can also use a time-dependent prior for time-dependent emission searches \cite{Braun}. 

\begin{figure}[t]
\includegraphics[width=0.5\textwidth,height=0.45\textwidth]{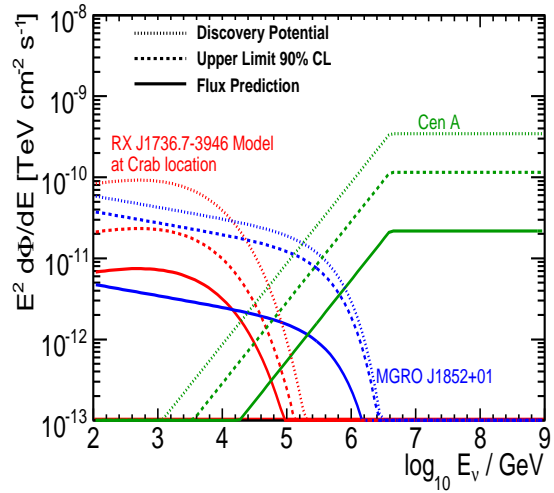} 
\caption{Differential flux for three theoretical models shown with the IceCube 40-string upper limit (90\% CL) and discovery potential in each case \cite{jon}.  Shown are the $\nu_{\mu}$ predictions of  for SNR RX J1713.7-3946 but moved to the location of the Crab Nebula, for MGRO J1852+01, and for Cen A with the optimistic condition that protons have a spectral index $\alpha_p=3$ (references in order in \cite{Morlino:2009ci}).}\label{fig:models}
\end{figure}

The search for the 40-string configuration \cite{jon} has been conducted on a data sample of 36,900 events collected during 375.5 days of livetime: 14,121 from the northern sky, mostly muons induced by upgoing atmospheric neutrinos that are filtered by the Earth and 22,779 from the southern sky, mostly high energy downgoing atmospheric muons.  
Since downgoing muons are about 5 orders of magnitude more numerous than atmospheric neutrinos, a zenith-dependent energy proxy cut has been implemented to reduce their number to the level of the atmospheric neutrino background from the other hemisphere. This region is then sensitive to neutrino sources with harder spectra in the southern sky 
compared to the other hemisphere. As a result, in the northern sky this search is sensitive in the TeV-PeV region and for the southern sky in the PeV-EeV one.
The sensitivity is at least a factor of 2 better than previous searches (depending on declination), with 90\% c.l. muon neutrino flux upper limits between $E^{2} dN/dE \sim 2 -- 200 \times 10^{-12} \mathrm{TeV \, cm^{-2} \, s^{-1}}$ in the northern sky and between $3 -- 700 \times 10^{-12} \mathrm{TeV \, cm^{-2}\, s^{-1}}$ in the southern sky. A comparison with some models is shown in Fig.~\ref{fig:models}.

 \begin{figure}[ht]
 \includegraphics[width=3.2in]{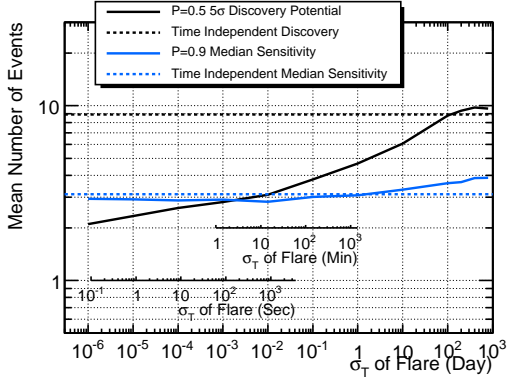} 
 \caption{The 50\% 5$\sigma$ discovery potential and 90\% median sensitivity versus flare duration (std of a Gaussian representing the flare) in terms of the mean number of events for a fixed source at $+16^{o}$ declination for the ``untriggered'' time-dependent search and for the time-independent search (preliminary). Flares with a duration of less than 100 days, or a FWHM of less than roughly half the total livetime, have a better discovery potential than the steady search \cite{mike}.}
 \label{timedep}
\end{figure}

\begin{figure}[t]
\includegraphics[width=0.45\textwidth,height=0.4\textwidth]{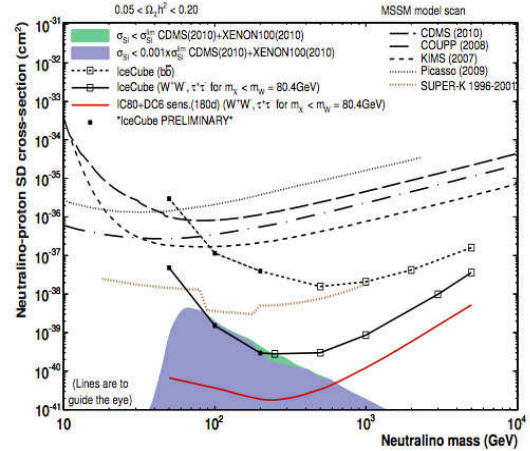}
\caption{90\% c.l. limits on the spin dependent p-neutralino cross section in the Sun vs the neutralino mass for the 22 string configuration and the AMANDA result for two indicated annihilation channels. The darkened regions represent MSSM parameter scans currently allowed by direct detection experiments. All references for other experiments and the description of IceCube analysis are in \cite{wimps}.}\label{fig:dm}
\end{figure}

When a time dependent term is added to the LH, this method is extensible to search for flares from sources like AGNs, GRBs and Soft-Gamma-ray Repeaters.
Time-integrated searches are less sensitive to flares because they are affected by a larger background of atmospheric neutrinos and muons 
that can be reduced by the time constraint.
In \cite{mike}  the parameter space of direction, energy and time has been scanned looking for clusters of high energy events in time and position with a method described in \cite{Braun}.
This so-called ``untriggered'' search covers any possible time-dependent emission from these sources not correlated to any observation using other astrophysical messengers such as photons. In the LH of the signal two additional nuisance parameters are added with respect to the time-independent search: the sigma and the mean of a flare assumed to have a gaussian shape.
In Fig.~\ref{timedep} it can be seen that for a fixed direction and for time emissions at the scale of 1 sec this search has a discovery potential 
about a factor of 4 better than time-independent searches. Also ``triggered'' searches by multi-wavelength information on flares from blazars and Soft Gamma Ray repeaters have been performed using lightcurves from continuously monitoring satellites such as Fermi-LAT and SWIFT and flares detected by Imaging Cherenkov Telescopes with shorter time-scale monitoring. When continuous lightcurves are available they can be used as priors in the LH method under the assumption that the neutrino emission is correlated to the one observed in photons. The definition of the flare duration is obtained maximizing the discovery potential
having added in the method as nuisance parameter the threshold of the emission. 
The triggered flare corresponding to the highest significance is from the blazar PKS 1502+106. It was seen by Fermi-LAT in gamma-rays and SWIFT in soft X-rays 
and optical, but not in hard X-rays. The probability after trials that this is due to a fluctuation of the background is 29\%. Each of these results are compatible with a fluctuation of the background.  

With the same method, we looked for neutrinos emitted by binary systems with period determined from optical observations.
We considered a catalogue of 7 selected binary systems mostly believed to be micro-quasars.
The most significant source in the catalogue is Cyg X-3 with a 1.8\% probability after trials ($2.1 \sigma$ one-sided), compatible
with fluctuations of the background. The best fit of the phase is 0.82, close to the superior conjunction of the system at phase = 0 and close to the peak of Fermi-LAT observations
in the gamma-rays.

\section{SEARCH FOR DARK MATTER}

\begin{figure}[t]
\includegraphics[width=0.45\textwidth,height=0.4\textwidth]{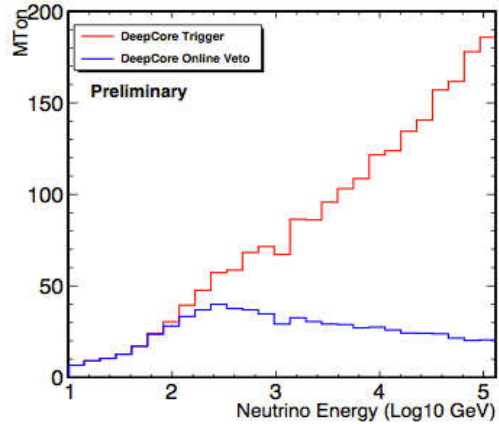}
\caption{The expected DeepCore muon effective volume after application of the SMT3 trigger (upper line) and after application of the online filter running at the South Pole (lower line) \protect\cite{deepcorepaper}. In this plot from simulation, neutrinos have been forced to interact in the DeepCore volume, and the muon neutrino interaction vertex can contribute hits and play a role in the triggering and filtering. High energy muon neutrinos that interact and produce a muon outside of the DeepCore volume are typically removed by the filter, causing the lower line in the plot to decrease at $E_{\nu} > 300$~GeV. From the DeepCore perspective, such events are indistinguishable from cosmic ray background.}\label{fig:veto}
\end{figure}

Neutrino telescopes indirectly search for the presence of dark matter gravitationally trapped inside celestial bodies like the Sun or at the Galactic Centre.
Over billions of years, a sufficiently large number of WIMPs can accumulate in the Sun's core to allow for their efficient annihilation. Such annihilations produce a wide range of particles, most of which are quickly absorbed into the solar medium. Neutrinos, on the other hand, may escape the Sun and be detected in experiments on the Earth. Since the Sun is rich in hydrogen, the limits set by IceCube for the spin-dependent cross section of the WIMP-nucleon interaction are competitive with respect to direct detection experiments (see Fig.~\ref{fig:dm}). While the limit begins to touch the region of the MSSM not excluded by direct detection experiments such as CDMS, COUPP and KIMS (see references in \cite{wimps}), IceCube with DeepCore in 1800 days of livetime will probe an interesting fraction of the parameter space (red dotted line).
The presence of DeepCore improves the detection capabilities at energies between 10-100 GeV by up to an order of magnitude with respect to IceCube 
standalone. DeepCore fiducial volume can be used to detect contained neutrino interactions and the external IceCube strings can be used as a veto to
reject downgoing atmospheric muons. Fig.~\ref{fig:veto} shows the effective volume of DeepCore at trigger (SMT3) level and online filter level. The efficiency for background rejection of the veto at filter level for the 59-string run is about $8 \times 10^{-3}$ and corresponds to a rate of atmospheric neutrinos of $5 \times 10^{-3}$~Hz while atmospheric muons have a rate of 7~Hz. Further reduction of background can be obtained at analysis level and with larger configurations (79 and 86 strings). The dedicated low energy trigger is similar to the SMT8 described in Sec.~\ref{sec1} but it requires the coincidence of only 3 DOMs (SMT3) in  a 2.5 $\mu$s time window.
By using the veto, searches such as dark matter, galactic sources with cut-offs at $1--10$~TeV, neutrinos from interactions of CRs in the galactic plane and neutrino
oscillations become very promising. Moreover the detection of neutrinos from SN collapse searches can also be enhanced.

\end{document}